\documentclass[conference]{IEEEtran}
\IEEEoverridecommandlockouts
\usepackage{cite}
\usepackage{amsmath,amssymb,amsfonts}
\usepackage{algorithmic}
\usepackage{hyperref}
\usepackage{graphicx}
\usepackage{multirow}
\usepackage{bbding}
\usepackage{textcomp}
\usepackage{makecell}
\usepackage{array}
\usepackage{booktabs}
\usepackage{pgfmath}
\usepackage{utfsym}
\usepackage{fontawesome}

\usepackage{cite}
\usepackage{algorithm}
\usepackage[numbers,sort&compress]{natbib}
\usepackage{comment}
\usepackage{subfigure}
\usepackage{arabtex}
\usepackage{adjustbox}
\usepackage{caption}
\usepackage{subcaption}

\usepackage{multicol}
\usepackage{color, colortbl}
\usepackage{xcolor}
\colorlet{eng}{blue!10}
\colorlet{cmn}{teal!10}
\colorlet{multi}{yellow!10}
\colorlet{euro}{orange!10}

\usepackage{setspace}

\usepackage{pifont} % include some cool symbols
\definecolor{bsRed}{rgb}{0.95, 0.0, 0.0}

\def\BibTeX{{\rm B\kern-.05em{\sc i\kern-.025em b}\kern-.08em
    T\kern-.1667em\lower.7ex\hbox{E}\kern-.125emX}}
\begin{document}

% \title{Hierarchy Content-Aware Audio-visual Temporal Forgery Localization\\
% {\footnotesize \textsuperscript{*}Note: Sub-titles are not captured for https://ieeexplore.ieee.org  and

% \title{MCMamba: An Empirical Study of Mamba for Multichannel Speech Enhancement 
% }
\title{Leveraging Joint Spectral and Spatial Learning with MAMBA for Multichannel Speech Enhancement
}

\author{
Wenze Ren$^1$, Haibin Wu$^1$, Yi-Cheng Lin$^1$, Xuanjun Chen$^1$, Rong Chao$^{24}$, Kuo-Hsuan Hung$^{34}$\\ You-Jin Li$^{14}$, Wen-Yuan Ting$^4$, Hsin-Min Wang$^4$, Yu Tsao$^4$ 
\\
\textit{$^1$Graduate Institute of
Communication Engineering, National Taiwan University}\\    \textit{$^2$Department of Computer Science
and Information Engineering, National Taiwan University}\\
\textit{$^3$Department of Biomedical Engineering, National Taiwan University\ }
\textit{$^4$Academia Sinica, Taiwan}\\

% \textit{$^1$Graduate Institute of
% Communication Engineering, National Taiwan University, Taipei, Taiwan}\\    \textit{$^2$Department of Computer Science
% and Information Engineering, National Taiwan University, Taipei, Taiwan}\\
% \textit{$^3$Department of Biomedical Engineering, National Taiwan University, Taipei, Taiwan}\\
% \textit{$^4$Research Center for Information Technology Innovation, Academia Sinica, Taipei, Taiwan}\\
% \textit{$^5$Institute of Information Science, Academia Sinica, Taipei, Taiwan}\\

% \textit{$^1$National Taiwan University, Taipei, Taiwan}\\
% \textit{$^2$Academia Sinica, Taipei, Taiwan}\\

\texttt{r11942166@ntu.edu.tw, yu.tsao@citi.sinica.edu.tw} 
}
\maketitle

\begin{abstract}
In multichannel speech enhancement, effectively capturing spatial and spectral information across different microphones is crucial for noise reduction. Traditional methods, such as CNN or LSTM, attempt to model the temporal dynamics of full-band and sub-band spectral and spatial features. However, these approaches face limitations in fully modeling complex temporal dependencies, especially in dynamic acoustic environments. To overcome these challenges, we modify the current advanced model McNet by introducing an improved version of Mamba, a state-space model, and further propose MCMamba. MCMamba has been completely reengineered to integrate full-band and narrow-band spatial information with sub-band and full-band spectral features, providing a more comprehensive approach to modeling spatial and spectral information. Our experimental results demonstrate that MCMamba significantly improves the modeling of spatial and spectral features in multichannel speech enhancement, outperforming McNet and achieving state-of-the-art performance on the CHiME-3 dataset. Additionally, we find that Mamba performs exceptionally well in modeling spectral information.
\end{abstract}


\section{Introduction}
\label{sec:intro}

Multichannel speech enhancement (SE) is a critical and challenging research topic in speech signal processing. 
In addition to the spectral information from each channel, the spatial information captured by microphone arrays is an important cue for multichannel SE. 
Compared to single channel methods \cite{wang2018supervisedspeechseparationbased, choi2018phase, Hao_2021, hu2020dccrn, wu23b_interspeech,lu13_interspeech}, multichannel techniques have the potential to significantly enhance speech quality by effectively incorporating both spatial and spectral information.
Spatial information refers to the difference in sound signals captured by multiple microphones, which helps determine the location of the sound source. 
This enables the SE systems to distinguish between speech and noise based on direction. 
Spectral information, on the other hand, refers to the frequency content of the signal, helping separate speech from noise by analyzing different frequency bands. 
Combining both spatial and spectral data improves the performance of SE.

Recently, there are models specifically designed to process spatial and spectral information across multiple channels. 
Notably, this advancement is exemplified by mechanisms such as the channel attention approach introduced in \cite{tolooshams2020channel}.
In addition, specific neural networks are designed to specifically target the spatial distinctions between narrow-band speech and noise \cite{li2019narrow, 8937218}, further optimizing spatial information processing. 
The spectral domain is another crucial feature for differentiating clean speech from noise. 
Researchers have demonstrated that both full-band and sub-band regions contain valuable spectral information \cite{xiong22_interspeech}. 
The state-of-the-art performance is achieved by modeling the spatial information in narrow-band and full-band frequencies while effectively combining sub-band and full-band features 
 \cite{yang2023mcnet}.

Mamba \cite{gu2023mamba}, a recent state-space model with a selection mechanism, has demonstrated high performance in single-channel speech separation \cite{li2024spmamba,jiang2024dual} and SE \cite{chao2024investigation}. 
Building on this, SpatialNet introduces a variant of Mamba tailored for long-term streaming multichannel SE \cite{10570301}, particularly suited for scenarios involving stationary and moving speakers. 
However, SpatialNet focuses primarily on utilizing spatial information. 
Research exploring the joint use of equally important spatial and spectral features in state-space models for multichannel SE remains unexplored.

This study proposes a novel multichannel SE system based on Mamba, which aims to effectively capture spatial and spectral information. 
We develop causal and non-causal versions of the system to meet practical application requirements. 
The specially designed Mamba module is dedicated to modeling spatial and spectral characteristics. 
Extensive experiments show that on the CHiME-3 dataset, our method achieves \textbf{state-of-the-art} performance, outperforming McNet \cite{yang2023mcnet} across all SE metrics.
% To further explore the capabilities of MAMBA in spatial and spectral modelling, we conducted a comparative experiment in which LSTM and MAMBA were alternately used for spatial and spectral modelling. 
% Carefully ablation study in controlled settings reveals Mamba is particularly advantageous in spectral modeling. 
A careful ablation study in controlled settings reveals that Mamba is particularly advantageous in spectral modeling.

\section{Methodology}

\begin{figure}[!t]
    \centering
    %\hspace{-0.9cm}
    {\includegraphics[width=.5\textwidth]{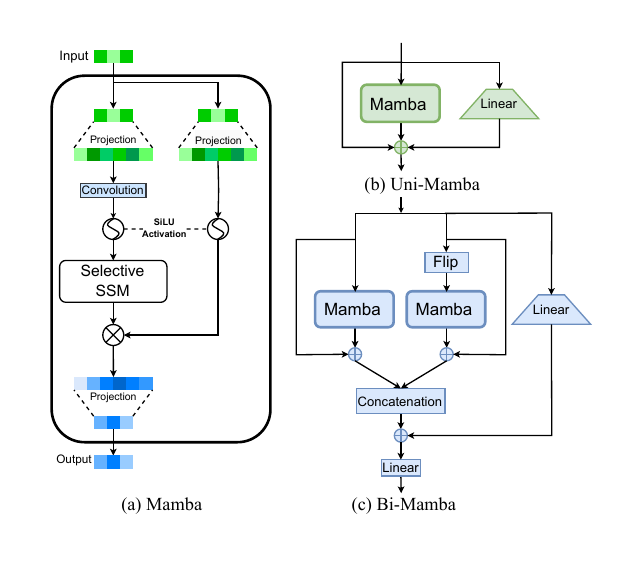}}
    \vspace{-1.0cm}
    \caption{(a)  The basic model architecture of Mamba, (b)  The proposed Uni-directional Mamba block, (c)  The proposed Bi-directional Mamba block.}
    \label{fig:Mamba_block}
    %\vspace{-0.3cm}
\end{figure}

This section first presents the Mamba framework and its two novel variants: Uni-Mamba and Bi-Mamba in Fig.~\ref{fig:Mamba_block}. 
Subsequently, we introduce the innovative MCMamba model, which builds upon and extends the capabilities of the Mamba and Bi-Mamba architectures. %Each model is tailored to address specific challenges in multichannel SE, utilizing advanced spatial and spectral processing techniques to optimize performance across different scenarios.

\subsection{Core Mamba Architecture}
Fig.~\ref{fig:Mamba_block}(a) illustrates the core architecture of Mamba. Mamba integrates the H3 architecture~\cite{fu2022hungry} and a gated multilayer perceptron block
into a stacked structure. Its main components include three linear layers, a deep convolutional network, and a selective State-Space Model (SSM). Given an input $X$ and an output $Y$, the system can be described by the following mathematical formulation:
\[
{Y}' \leftarrow \text{SSM}\left(\text{SiLU}\left(\text{Conv}\left(\text{Linear}(X)\right)\right)\right) 
\odot \text{SiLU}\left(\text{Linear}(X)\right),
\]
\[
{Y} \leftarrow \text{Linear}{(Y')}.
\]

The selective SSM, which is the core processing unit in Mamba, maps an input $x$ to an output $y$ through a higher-dimensional latent state $h$, described by the following equations:
\[
h_n = \bar{A}h_{n-1} + \bar{B}x_n,
\]
\[
y_n = Ch_n,
\]
where $\bar{A}$ and $\bar{B}$ represent discretized state matrices. The discretization process transforms continuous parameters $(\Delta, A, B)$ into their discrete counterparts $(\bar{A}, \bar{B})$, enabling the model to process discrete-time audio signals effectively\cite{gu2023mamba}.

The Structured SSM~\cite{gu2021efficiently} offers a compelling alternative to convolutional neural networks (CNNs) and recurrent neural networks (RNNs) for handling long-range dependencies in sequential data~\cite{gu2022parameterization}. Building on this foundation, the Mamba model incorporates a novel input selection mechanism, effectively merging the strengths of CNNs and RNNs, significantly enhancing its capacity to model discrete data within the SSM framework. This mechanism allows the model to dynamically parameterize SSM components based on the input, selectively focusing on or ignoring specific parts of the input sequence. Consequently, Mamba excels at efficiently processing complex and extended sequences.
%%%%%%%%%%%%%%%%%%%%%%%%%%%%%%%%%%%%%%%%%%%%%%%%%%%%%%%%%%%%%%%%%%%%%%%%%%%%%%
\begin{figure*}[!htbp]
    \centering
    \vspace{-0.5cm}
    \includegraphics[width=1\textwidth]{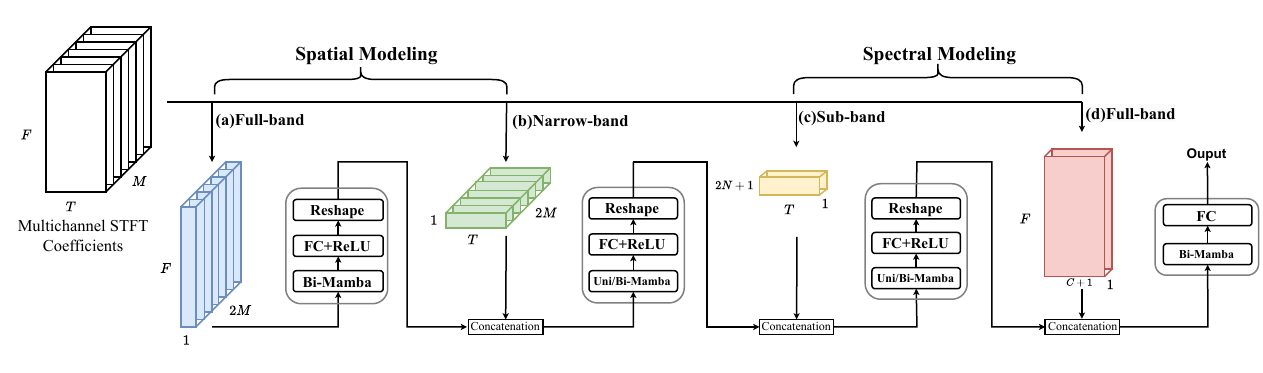}
    \vspace{-0.1cm}
    \caption{Diagram of the proposed MCMamba architecture.}
    \label{fig:Mcmamba}
\end{figure*}
%%%%%%%%%%%%%%%%%%%%%%%%%%%%%%%%%%%%%%%%%%%%%%%%%%%%%%%%%%%%%%%%%%%%%%%%%%%%%%
\subsection{Uni-directional and bi-directional Mamba}
To address the distinct requirements of real-time and offline multichannel SE, we propose two variants of the Mamba model: Uni-Mamba in Fig.~\ref{fig:Mamba_block}(b) and Bi-Mamba in Fig.~\ref{fig:Mamba_block}(c). 

Uni-Mamba is a causal model specifically designed for real-time applications, where only the past and present information is processed. 
This characteristic makes Uni-Mamba particularly well-suited for real-time multichannel SE tasks, ensuring efficient performance under strict latency constraints.
A key innovation is to \textbf{modify the traditional residual connection}. 
Unlike SPmamba \cite{li2024spmamba} and SEmamba \cite{chao2024investigation}, which combine Mamba with a traditional residual structure, we add an additional learnable linear residual path. This provides the model with an additional representation that complements other paths. The linear residual connection of Uni-Mamba has an input with the same dimension as the output.
This modification introduces greater flexibility and enhances the model's capacity to capture more complex representations and dependencies within the speech data, ultimately improving the performance in multichannel SE as shown in the experiments.

Bi-Mamba, in contrast, is a non-causal model designed for offline processing. 
Like Uni-Mamba, the strategy it adopts is to project the input of the Mamba module to a larger dimension (twice the original) and then form a linear residual connection.
The architecture utilizes two Mamba blocks: one processes the input in its original direction, while the other handles the flipped input. 
The outputs from both directions are combined and refined through a final linear layer. 
By leveraging past and future information, Bi-Mamba performs better in SE tasks.

\subsection{Proposed Method: MCMamba}
In this work, we present MCMamba in Fig.~\ref{fig:Mcmamba}, a multichannel SE network specifically designed to process temporal and spectral cues through four dedicated modules: full-band spatial, narrow-band spatial, sub-band spectral, and full-band spectral. 
While inspired by the McNet framework, MCMamba introduces a novel variant of Mamba to further enhance its capabilities. 
Each module is carefully structured to progressively refine the features extracted from the preceding stage, ensuring that essential information is preserved throughout the process. 
MCMamba offers flexibility based on latency requirements, utilizing Bi-Mamba for non-causal (offline) processing and Uni-Mamba for causal (real-time) processing.

\subsubsection{spatial modeling}
Spatial feature modeling in MCMamba comprises two distinct modules: the full-band spatial module (Fig.~\ref{fig:Mcmamba}(a)) and the narrow-band spatial module (Fig.~\ref{fig:Mcmamba}(b)). 
Each module is responsible for extracting spatial features across different frequency ranges in multichannel speech, ensuring comprehensive spatial representation across the spectrum.
The spatial characteristics across the entire frequency band are captured by processing the multichannel Short-Time Fourier Transform (STFT) coefficients along the frequency axis, effectively modeling the spatial correlations across all frequencies. 
% The input comprises both the real(Re) and imaginary(Im) components of the multichannel signal, represented by the following formula:
The spatial information of the input signal at each time-frequency point $(t, f)$ is represented by concatenating the real (Re) and imaginary (Im) parts of STFT coefficients from all $M$ channels:
\begin{equation}
    x^1(t, f) = \begin{aligned}[t]
&[ \text{Re}(X_1(t, f)), \text{Im}(X_1(t, f)), \dots, \\
&\text{Re}(X_M(t, f)), \text{Im}(X_M(t, f))]^T.
\end{aligned}
\end{equation}
% \[
% x(t, f) = \begin{aligned}[t]
% &[ \text{Re}(X_1(t, f)), \text{Im}(X_1(t, f)), \dots, \\
% &\text{Re}(X_M(t, f)), \text{Im}(X_M(t, f))]^T
% \end{aligned}
% \]
%%%%%%%%%%%%%%%%%%%%%%%%%%%%%%%%%%%%%%%%%%%%%%%%%%%%%%%%%%%%%%%%%%%%%%%%%%%
This allows learning of spatial cue correlations across frequencies, such as the inter-channel phase difference (IPD).
% where $M$ denotes the microphone number. 
% Spatial cues, such as the inter-channel phase difference (IPD), are then extracted to capture the spatial relationships between the channels:
% \begin{equation}
%     \text{IPD}(f) = \angle X_i(t, f) - \angle X_j(t, f)
% \end{equation}
% % \[
% % \text{IPD}(f) = \angle X_i(t, f) - \angle X_j(t, f)
% % \]
%%%%%%%%%%%%%%%%%%%%%%%%%%%%%%%%%%%%%%%%%%%%%%%%%%%%%%%%%%%%%%%%%%%%%%%%%%%
The component does not capture temporal dependencies (always in a causal manner); therefore, the Bi-Mamba module extracts the full-band spatial features. 
Fully connected (FC) layers with ReLU activation are used to project the output to the desired dimension, after which the data is reshaped for further spatial processing.
%%%%%%%%%%%%%%%%%%%%%%%%%%%%%%%%%%%%%%%%%%%%%%%%%%%%%%%%%%%%%%%%%%%%%%
% The narrow-band spatial module (Fig.~\ref{fig:Mcmamba} (b)) captures the spatial information within each frequency band and models the evolution of these spatial cues. %how these spatial cues evolve. 
% The narrow-band process below will be executed several times in parallel for each narrow band. 

The narrow-band spatial module (Fig.~\ref{fig:Mcmamba}(b)) captures the spatial information within each frequency band and models its temporal evolution. %how these spatial cues evolve. 
The narrow-band process below will be executed several times in parallel for each narrow band. 
% The output from the full-band spatial module and the original noisy multichannel signal are passed to the narrow-band module in a cascade manner across the channel dimension. 
% The time-dependent representation is formulated as follows:
The concatenation of the original multichannel STFT coefficients and the output from the full-band spatial module are passed to the narrow-band module. Different frequencies are processed independently but share the same network parameters. The temporal sequence at one frequency can be formulated as follows:
%%%%%%%%%%%%%%%%%%%%%%%%%%%%%%%%%%%%%%%%%%%%%%%%%%%%%%%%%%%%%%%%%%%%%%%%%%%
\begin{equation}
    x^2(f) = [x(1, f), x(2, f), \dots, x(T, f)]. 
\end{equation}
% \[
% X(f) = [x(1, f), x(2, f), \dots, x(T, f)] 
% \]
By employing either the Bi-Mamba (non-causal) or Uni-Mamba (causal) module, followed by a FC layer to refine spatial characteristics, MCMamba effectively divides spatial modeling into full-band and narrow-band components. 
This approach ensures that frequency and time domain correlations across channels are fully leveraged, resulting in a more accurate separation of speech from noise.

\subsubsection{spectral modeling}
Spectrum modeling in MCMamba is carried out through the sub-band and full-band modules, each designed to process the frequency and spectral information of the reference channel. 
The sub-band spectral module (Fig.~\ref{fig:Mcmamba}(c)) is tasked with capturing local spectral patterns and signal smoothness, both of which are critical for differentiating speech from noise. 
This module processes the STFT amplitude of the reference channel, namely $|X_r(t, f)|$.
%\begin{equation}
%    X_r(t, f) = |X_r(t, f)|
%\end{equation}
% \[
% X_r(t, f) = |X_r(t, f)|
% \]
The data from adjacent frequencies are interconnected to utilize spectral information more effectively:
\begin{equation}
    x^3(t, f) = [|X_r(t, f-N)|, \dots, |X_r(t, f+N)|],
\end{equation}
% \[
% x_4(t, f) = [|X_r(t, f-N)|, \dots, |X_r(t, f+N)|]
% \]
where $N$ is the number of neighbors considered.
Depending on whether real-time or offline processing is required, either Bi-Mamba or Uni-Mamba is employed to capture the partial spectral dynamics over time, followed by a FC layer with ReLU activation for further refinement.

The full-band spectral module (Fig.~\ref{fig:Mcmamba}(d)) integrates information across the entire frequency range to model a broad-spectrum pattern. 
It operates along the frequency axis, combining multiple consecutive frames to capture temporal dependencies within the broader spectral context. 
The model gains access to more comprehensive information by connecting these context frames:
\begin{equation}
    x^4(t, f) = [|X_r(t-C, f)|, \dots, |X_r(t, f)|],
    \label{eq:(d)}
\end{equation}
where $C$ is the number of context frames considered.
% \[
% x(t, f) = [|X_r(t-C, f)|, \dots, |X_r(t, f)|]
% \]
By utilizing both sub-band and full-band modules, MCMamba effectively captures local and global spectral cues, ensuring that spectral modeling complements spatial information to enhance the overall accuracy of speech reconstruction.

\section{Experiments}
\begin{table}[h]
\centering
\small
\setlength{\tabcolsep}{1.5pt}
\captionsetup{font=footnotesize}
\caption{Speech enhancement performance comparison. $^\dagger$Cited from other original papers, $^*$Cited from McNet. Some papers don't provide causal results.}
\label{table:speech_enhancement}
\begin{tabular}{l c c c c}
\toprule
\textbf{Method} & \textbf{NB-PESQ} & \textbf{WB-PESQ} & \textbf{STOI} & \textbf{SDR} \\
\midrule
\multicolumn{5}{c}{\textbf{Performance of Non-Causal (Offline) speech enhancement}} \\
\midrule
Noisy & 1.82 & 1.27 & 87.0 & 7.5 \\
\midrule
MNMF Beamforming$^\dagger$\cite{8673623} & -- & -- & 94.0 & 16.2 \\
Oracle MVDR$^*$ & 2.49 & 1.94 & 97.0 & 17.3 \\
CA Dense U-net$^\dagger$\cite{tolooshams2020channel} & -- & 2.44 & -- & 18.6 \\
Narrow-band Net$^*$\cite{li2019narrow} & 2.74 & 2.13 & 95.0 & 16.6 \\
FT-JNF$^*$\cite{Tesch_2023} & 3.17 & 2.48 & 96.2 & 17.7 \\
McNet$^*$\cite{yang2023mcnet} & 3.38 & 2.73 & 97.6 & 19.6 \\
MVDR-Embedded U-Net$^\dagger$\cite{10448366} & -- & 2.64 & 97.4 & 20.6 \\
\midrule
MCMamba & \textbf{3.49} & \textbf{2.98} & \textbf{98.2} & \textbf{20.8} \\
\midrule
\midrule
\multicolumn{5}{c}{\textbf{Performance of Causal (Online) speech enhancement}} \\
\midrule
Noisy & 1.82 & 1.27 & 87.0 & 7.5 \\
\midrule
Narrow-band Net$^*$\cite{li2019narrow} & 2.70 & 2.15 & 94.7 & 16.0 \\
FT-JNF$^*$\cite{Tesch_2023} & 2.80 & 2.23 & 95.4 & 16.9 \\
McNet$^*$\cite{yang2023mcnet} & 3.29 & 2.67 & 97.2 & 19.0 \\
\midrule
MCMamba & \textbf{3.41} & \textbf{2.89} & \textbf{97.6} & \textbf{19.8} \\
\bottomrule
\end{tabular}
\end{table}
\begin{table}[h]
\centering
\small
\setlength{\tabcolsep}{2pt}
\setlength\tabcolsep{4pt}
\captionsetup{font=footnotesize}
\caption{Ablation study on modules for spatial and spectral modeling, where the spatial features of the narrow band in Fig.~\ref{fig:Mcmamba}(b) and the spectral features of the sub-band Fig.~\ref{fig:Mcmamba}(c) are time-dependent.}
\label{table:spatial_spectral_comparison}
\begin{tabular}{cc|cccc}
\toprule
\textbf{Spatial} & \textbf{Spectral} & \textbf{NB-PESQ} & \textbf{WB-PESQ} & \textbf{STOI} & \textbf{SDR} \\
\midrule
LSTM & LSTM & 3.27 & 2.72 & 97.46 & 19.30 \\
Mamba & LSTM & 3.46 & 2.80 & 97.60 & 19.22 \\
LSTM & Mamba & \textbf{3.52} & \textbf{3.02} & \textbf{98.10} & \textbf{20.50} \\
\bottomrule
\end{tabular}
\end{table}

\subsection{Experimental setup}

\subsubsection{Dataset}
We used the simulated dataset from the 3rd CHiME Challenge, comprising 7,138, 1,640, and 1,320 utterances for training, development, and testing. 
The data, sampled at 16 kHz using a six-microphone tablet, includes multichannel noise from four environments: bus, cafe, pedestrian area, and street intersection. 
The official CHiME-3 Matlab script was used to generate the dataset. 
We dynamically generated samples with random noise clips and varied SNR between -5 and 10 dB to enhance training. 
Speech signals were simulated by adding delays to single-channel utterances without reverberation. 
A Hann window of 512 samples with a 256-sample step size was applied for processing.

\subsubsection{Model configurations} 
The output sizes of the Mamba blocks (either Uni- or Bi-) for full-band spatial, narrow-band spatial, sub-band spectral, and full-band spectral modeling are 64, 64, 64, and 2. 
The corresponding hidden cell sizes are 128, 256, 384, and 128, respectively. 
The sub-band spectral model uses 6 adjacent frequency bands ($N=3$), while the full-band spectral model uses 5 context frames ($C=5$).

\subsubsection{Training setting}
% The model was trained for 500 epochs until convergence, 
The max epoch is 500, with the best model selected based on the validation set, with an initial learning rate of 0.001 and a decay rate of 0.992 per epoch, using the Adam optimizer. 
The 5th channel was selected as the reference channel, and the performance was evaluated using NB-PESQ and WB-PESQ \cite{941023}, STOI \cite{5713237}, and SDR \cite{1643671}.

\subsubsection{Baseline methods}
We compare our method with the following multichannel SE techniques: 
MNMF Beamforming \cite{8673623}, which employs multichannel non-negative matrix factorization (MNMF) followed by beamforming for SE; 
Oracle MVDR, which utilizes the actual noise spatial covariance matrix for enhancement; 
Narrow-band Net \cite{li2019narrow}, which uses two LSTMs to capture narrow-band spatial information; 
FT-JNF \cite{Tesch_2023}, an improved version of Narrow-band Net that more effectively leverages full-frequency information; 
McNet \cite{yang2023mcnet} (previous SOTA), a fusion network that incorporates LSTMs to model both spatial and spectral information; 
and MVDR-Embedding U-Net \cite{10448366}, which integrates the minimum variance distortionless response (MVDR) module into the U-Net architecture, combining beamforming with neural networks for enhanced speech quality.

\subsection{Experimental results}
\subsubsection{Results and analysis}
In Table~\ref{table:speech_enhancement}, we fist compare the SE performance of various non-causal models. 
Evidently, traditional signal processing techniques, such as the MVDR-Embedded U-Net combined with deep learning methods, often yield superior outcomes. 
Notably, FT-JNF, which effectively leverages spatial information, outperforms Narrow-band Net. 
McNet, which integrates spatial and spectral features, underscores the critical importance of utilizing spectral information in multichannel SE. 
Our proposed MCMamba model achieves the \textbf{best} performance, demonstrating the strength of Bi-Mamba in modeling spatial and spectral information across multiple channels.

Given the significance of real-time SE, we also compare the performance of causal models. 
Again, MCMamba delivers highly competitive results, \textbf{surpassing even all the baseline non-causal models}. 
This highlights the effectiveness of our Mamba module in efficiently exploiting spatial information across different bandwidths and spectral cues across frequency bands. 
Notably, despite the absence of future information, MCMamba achieves excellent enhancement performance, affirming the robustness of the proposed approach.

\subsubsection{Ablation study}
% To further explore the capability of MAMBA in modeling spatial and spectral information, we conducted a series of ablation studies where we alternated between using LSTM and MAMBA for the spatial and spectral blocks. The goal of these experiments was to evaluate the effectiveness of MAMBA in comparison to LSTM, which is commonly used for temporal modeling.

% In our experimental setup, we examined the following configurations:
% \begin{itemize}
%     \item \textbf{LSTM + LSTM}: Both the spatial and spectral blocks utilize LSTM for modeling.
%     \item \textbf{MAMBA + LSTM}: MAMBA is used for spatial modeling, while LSTM is used for spectral modeling.
%     \item \textbf{LSTM + MAMBA}: LSTM is used for spatial modeling, while MAMBA is applied to spectral modeling.
% \end{itemize}
Our ablation study further demonstrates Mamba's capabilities in spatial and spectral modeling. 
% We used LSTM, which excels at temporal modeling, to alternate with the Mamba blocks we designed for spatial and frequency modeling. 
%\textcolor{red}{We used LSTM, known for their exceptional temporal modeling, to alternate with the Mamba blocks we designed for spatial and frequency modeling.}
Since the LSTM architecture is known for exceptional temporal modeling, we compared its spatial and spectral modeling capabilities with that of the Mamba blocks we designed.
Note that the models in Table~\ref{table:spatial_spectral_comparison} are causal, i.e., Uni-Mamba or LSTM is used to model temporal correlation.
Bi-Mamba or BLSTM is only used to model frequency correlation.
All three models in Table~\ref{table:spatial_spectral_comparison} are with similar parameter numbers.
% Table \ref{table:spatial_spectral_comparison} shows the results of the comparison experiments. 
It serves as the baseline when modeling spatial and spectral features using LSTM. 
When Mamba is introduced for spectral or spatial modeling, all metrics improves dramatically.
It is worth noting that when using Mamba for spectral modeling, the performance is significantly improved, especially for NB-PESQ and WB-PESQ, which are highly sensitive to spectral information. 
% This demonstrates the advantages of Mamba over LSTM in modeling spectral information, especially its more remarkable ability to capture complex spectral dynamics.
This demonstrates Mamba's advantages over LSTM in modeling spectral information, especially its superior ability to capture complex spectral dynamics.

\section{Conclusion}
This paper proposes MCMamba, a novel model designed for multichannel speech enhancement, leveraging spatial and spectral information through Uni-Mamba and Bi-Mamba models for causal and non-causal processing. 
MCMamba achieves state-of-the-art performance on the CHiME-3 dataset.
Our experiments also show that the Mamba architecture significantly improves spectral modeling over traditional methods. 
% This confirms MCMamba as a robust solution, achieving state-of-the-art performance in dynamic acoustic environments with effective spectral and spatial joint modeling.

% \newpage

\setstretch{1}
\bibliographystyle{IEEEtran}
\bibliography{IEEEabrv,refs}

\begin{thebibliography}{10}
\providecommand{\url}[1]{#1}
\csname url@samestyle\endcsname
\providecommand{\newblock}{\relax}
\providecommand{\bibinfo}[2]{#2}
\providecommand{\BIBentrySTDinterwordspacing}{\spaceskip=0pt\relax}
\providecommand{\BIBentryALTinterwordstretchfactor}{4}
\providecommand{\BIBentryALTinterwordspacing}{\spaceskip=\fontdimen2\font plus
\BIBentryALTinterwordstretchfactor\fontdimen3\font minus \fontdimen4\font\relax}
\providecommand{\BIBforeignlanguage}[2]{{%
\expandafter\ifx\csname l@#1\endcsname\relax
\typeout{** WARNING: IEEEtran.bst: No hyphenation pattern has been}%
\typeout{** loaded for the language `#1'. Using the pattern for}%
\typeout{** the default language instead.}%
\else
\language=\csname l@#1\endcsname
\fi
#2}}
\providecommand{\BIBdecl}{\relax}
\BIBdecl

\bibitem{wang2018supervisedspeechseparationbased}
D.~Wang and J.~Chen, ``Supervised speech separation based on deep learning: An overview,'' \emph{IEEE/ACM transactions on audio, speech, and language processing}, vol.~26, no.~10, pp. 1702--1726, 2018.

\bibitem{choi2018phase}
H.-S. Choi, J.-H. Kim, J.~Huh, A.~Kim, J.-W. Ha, and K.~Lee, ``Phase-aware speech enhancement with deep complex u-net,'' in \emph{Proc. ICLR}, 2018.

\bibitem{Hao_2021}
X.~Hao, X.~Su, R.~Horaud, and X.~Li, ``Fullsubnet: A full-band and sub-band fusion model for real-time single-channel speech enhancement,'' in \emph{Proc. ICASSP}, 2021.

\bibitem{hu2020dccrn}
Y.~Hu, Y.~Liu, S.~Lv, M.~Xing, S.~Zhang, Y.~Fu, J.~Wu, B.~Zhang, and L.~Xie, ``{DCCRN}: Deep complex convolution recurrent network for phase-aware speech enhancement,'' \emph{arXiv preprint arXiv:2008.00264}, 2020.

\bibitem{wu23b_interspeech}
H.~Wu, K.~Tan, B.~Xu, A.~Kumar, and D.~Wong, ``Rethinking complex-valued deep neural networks for monaural speech enhancement,'' in \emph{Proc. Interspeech}, 2023, pp. 3889--3893.

\bibitem{lu13_interspeech}
X.~Lu, Y.~Tsao, S.~Matsuda, and C.~Hori, ``Speech enhancement based on deep denoising autoencoder,'' in \emph{Proc. Interspeech}, 2013, pp. 436--440.

\bibitem{tolooshams2020channel}
B.~Tolooshams, R.~Giri, A.~H. Song, U.~Isik, and A.~Krishnaswamy, ``Channel-attention dense {U-N}et for multichannel speech enhancement,'' in \emph{Proc. ICASSP}, 2020, pp. 836--840.

\bibitem{li2019narrow}
X.~Li and R.~Horaud, ``Narrow-band deep filtering for multichannel speech enhancement,'' \emph{arXiv preprint arXiv:1911.10791}, 2019.

\bibitem{8937218}
------, ``Multichannel speech enhancement based on time-frequency masking using subband long short-term memory,'' in \emph{Proc. WASPAA}, 2019, pp. 298--302.

\bibitem{xiong22_interspeech}
F.~Xiong, W.~Chen, P.~Wang, X.~Li, and J.~Feng, ``Spectro-temporal subnet for real-time monaural speech denoising and dereverberation,'' in \emph{Proc. Interspeech}, 2022.

\bibitem{yang2023mcnet}
Y.~Yang, C.~Quan, and X.~Li, ``Mc{N}et: Fuse multiple cues for multichannel speech enhancement,'' in \emph{Proc. ICASSP}, 2023, pp. 1--5.

\bibitem{gu2023mamba}
A.~Gu and T.~Dao, ``Mamba: Linear-time sequence modeling with selective state spaces,'' \emph{arXiv preprint arXiv:2312.00752}, 2023.

\bibitem{li2024spmamba}
K.~Li, G.~Chen, R.~Yang, and X.~Hu, ``Spmamba: State-space model is all you need in speech separation,'' \emph{arXiv preprint arXiv:2404.02063}, 2024.

\bibitem{jiang2024dual}
X.~Jiang, C.~Han, and N.~Mesgarani, ``Dual-path {M}amba: Short and long-term bidirectional selective structured state space models for speech separation,'' \emph{arXiv preprint arXiv:2403.18257}, 2024.

\bibitem{chao2024investigation}
R.~Chao, W.-H. Cheng, M.~La~Quatra, S.~M. Siniscalchi, C.-H.~H. Yang, S.-W. Fu, and Y.~Tsao, ``An investigation of incorporating mamba for speech enhancement,'' \emph{arXiv preprint arXiv:2405.06573}, 2024.

\bibitem{10570301}
C.~Quan and X.~Li, ``Multichannel long-term streaming neural speech enhancement for static and moving speakers,'' \emph{IEEE Signal Processing Letters}, vol.~31, pp. 2295--2299, 2024.

\bibitem{fu2022hungry}
D.~Y. Fu, T.~Dao, K.~K. Saab, A.~W. Thomas, A.~Rudra, and C.~R{\'e}, ``Hungry hungry hippos: Towards language modeling with state space models,'' \emph{arXiv preprint arXiv:2212.14052}, 2022.

\bibitem{gu2021efficiently}
A.~Gu, K.~Goel, and C.~R{\'e}, ``Efficiently modeling long sequences with structured state spaces,'' \emph{arXiv preprint arXiv:2111.00396}, 2021.

\bibitem{gu2022parameterization}
A.~Gu, K.~Goel, A.~Gupta, and C.~R{\'e}, ``On the parameterization and initialization of diagonal state space models,'' \emph{Advances in Neural Information Processing Systems}, vol.~35, pp. 35\,971--35\,983, 2022.

\bibitem{8673623}
K.~Shimada, Y.~Bando, M.~Mimura, K.~Itoyama, K.~Yoshii, and T.~Kawahara, ``Unsupervised speech enhancement based on multichannel nmf-informed beamforming for noise-robust automatic speech recognition,'' \emph{IEEE/ACM Transactions on Audio, Speech, and Language Processing}, vol.~27, no.~5, pp. 960--971, 2019.

\bibitem{Tesch_2023}
K.~Tesch and T.~Gerkmann, ``Insights into deep non-linear filters for improved multi-channel speech enhancement,'' \emph{IEEE/ACM Transactions on Audio, Speech, and Language Processing}, vol.~31, p. 563–575, 2023.

\bibitem{10448366}
C.-H. Lee, K.~Patel, C.~Yang, Y.~Shen, and H.~Jin, ``An {MVDR}-embedded {U-Net} beamformer for effective and robust multichannel speech enhancement,'' in \emph{Proc. ICASSP}, 2024, pp. 8541--8545.

\bibitem{941023}
A.~Rix, J.~Beerends, M.~Hollier, and A.~Hekstra, ``Perceptual evaluation of speech quality (pesq)-a new method for speech quality assessment of telephone networks and codecs,'' in \emph{Proc. ICASSP}, vol.~2, 2001, pp. 749--752 vol.2.

\bibitem{5713237}
C.~Taal, R.~Hendriks, R.~Heusdens, and J.~Jensen, ``An algorithm for intelligibility prediction of time–frequency weighted noisy speech,'' \emph{IEEE Transactions on Audio, Speech, and Language Processing}, vol.~19, no.~7, pp. 2125--2136, 2011.

\bibitem{1643671}
E.~Vincent, R.~Gribonval, and C.~Fevotte, ``Performance measurement in blind audio source separation,'' \emph{IEEE Transactions on Audio, Speech, and Language Processing}, vol.~14, no.~4, pp. 1462--1469, 2006.

\end{thebibliography}

\end{document}